\documentstyle[11pt]{article}

\catcode`\@=11 \@addtoreset{equation}{section}

\catcode`\@=12

\def\I{I \hskip -2.0mm I}

\begin{document}

\title{There might be superluminal particles in nature
\thanks{The paper has been published in Journal of Shaanxi Normal University (Natural Science
Edition),Vol.29,No.3}}
\author{Guang-jiong Ni \thanks{E-mail: gj\_ni@Yahoo.com}\\
\small\it{Department of Physics, Fudan University, Shanghai, 200433, P.R.China}\\
\small\it{Department of Physics, Portland State University, Portland, OR 97207 USA.}}
\date{}

\maketitle
\begin{abstract}
Based on experimental discovery that the mass-square of neutrino is negative, a
quantum equation for superluminal neutrino is proposed in comparison with Dirac
equation and Dirac equation with imaginary mass. A virtual particle may also be viewed
as superluminal one. Both the basic symmetry of space-time inversion and the maximum
violation of space-inversion symmetry are emphasized.
\vskip 4mm \noindent{\bf Key
words:} ordinary neutrinos; neutrino mass and mixing; non-standard-model neutrinos
\end{abstract}

\section{INTRODUCTION}\label{sec-1}
\hskip\parindent
Nothing can travel faster than light. Is this statement as true now
as it ever was? In 2000 there were two experiments showing superluminal propagation of
microwave $^{[1]}$ or laser pulse $^{[2]}$. Yet physicists believe that the law of
physics have remained intact $^{[3]}$. However, the experimental discovery of negative
mass square of neutrino in recent years $^{[4]}$, $$ E^2=c^2p^2+m^2c^4,\eqno{(1)} $$
$$ m^2(\nu _e)=-2.5\pm 3.3eV^2,\eqno{(2)} $$ though far from accurate, does strongly
hint that neutrino might be a particle moving faster than light. Actually, rewriting
(1) as $$ E^2=c^2p^2-m_s^2c^4\eqno{(3)} $$ and using the quantum relations
$E=\hbar\omega$ and $p=\hbar k$, one easily derives the kinematic relation for
superluminal particle ( also named as tachyon in literature ) as follows ($m_s$ is
called "proper mass"): $$
p=\tilde{m}u=\frac{m_su}{\sqrt{\frac{u^2}{c^2}-1}},\quad E=\tilde{m}c^2=%
\frac{m_sc^2}{\sqrt{\frac{u^2}{c^2}-1}}.\eqno{(4)} $$
Here the velocity of particle
$u$ is identified with the group velocity $u_g=\frac{d\omega }{dk}$ of wave. To derive
(3), a quantum equation for neutrino was proposed in Ref. [5,6] (see also [7]). In
this paper, we will compare three kinds of equation---Dirac equation and two
"superluminal" equations and discuss relevant problems.

\section{Dirac equation versus superluminal equation}
\hskip\parindent
The Dirac equation for fermion with rest mass $m_0$ reads
$$
i\hbar \frac{\partial}{\partial t}\psi = ic\hbar\vec{\alpha}%
\cdot\nabla\psi+\beta m_0c^2\psi\eqno{(5)} $$ where $\psi$ is a four-component spinor
wave function and $\alpha_i$ and $\beta$ are 4$\times$4 matrixes. In Dirac
representation, they are expressed as $$ \psi=\left(\begin{array}{c}\varphi \\
\chi\end{array}\right),\quad \alpha_i= \left(\begin{array}{cc}0 & \sigma_i \\ \sigma_i
& 0\end{array}\right),\quad \beta= \left(\begin{array}{cc}I & 0 \\ 0 &
-I\end{array}\right)\eqno{(6)} $$ where $\varphi$ and $\chi$ are two-component
spinors, $\sigma_i$ are Pauli matrixes. Now we perform a unitary transformation as $$
\psi\longrightarrow U\psi=\frac{1}{\sqrt{2}}\left(\begin{array}{cc}1 & 1 \\ 1 &
-1\end{array} \right)\psi=\psi^{\prime}=\left(\begin{array}{c}\xi \\
\eta\end{array}\right).\eqno{(7)} $$ Let the matrix $U$ act on Eq.(5) from the left,
due to noncommutativity between $U$ and $\alpha_i$ ( or $\beta$ ), we find Dirac
equation in "Weyl representation" with $$ \psi=\left(\begin{array}{c}\xi \\
\eta\end{array}\right),\quad \alpha_i^{(W)}= \left(\begin{array}{cc}\sigma_i&o\\
0&-\sigma_i \end{array}\right),\quad \beta^{(W)}= \left(\begin{array}{cc}0&I
\\I&0\end{array}\right)\eqno{(8)} $$ Note that, however, the representation
transformation leads to important change in physical interpretation. While $\varphi$
and $\chi$ in (6) represent the "hidden particle and antiparticle fields" in a
particle state$^{[8]}$, $\xi$ and $\eta$ in (8) characterize the "hidden left-handed
and right-handed rotating fields" in a particle with 100\% left-handed helicity
explicitly.

Now let's perform a unitary transformation on Dirac equation (5) in Weyl representation:
$$
\psi=\left(\begin{array}{c}\xi \\ \eta\end{array}\right)
\longrightarrow U_s \psi=\left(\begin{array}{cc}i & 0 \\ 0 & 1\end{array}\right)\psi
=\psi_s=\left(\begin{array}{c}\xi_s \\ \eta_s\end{array}\right).\eqno{(9)}
$$
After setting
$$m_s=-im_0,\eqno{(10)}$$
we obtain an equation with real proper mass $m_s$:
$$
\begin{array}{l}
i\hbar \frac \partial {\partial t}\xi_s =ic\hbar \vec{\sigma}\cdot \nabla \xi_s-m_sc^2\eta_s ,
\nonumber \\
i\hbar \frac \partial {\partial t}\eta_s =-ic\hbar \vec{\sigma}\cdot \nabla\eta_s +m_sc^2\xi_s.
\end{array}\eqno{(11)}
$$

A substitution of plane-wave solution $\xi_s\sim\eta_s\sim exp[i(px-Et)/\hbar]$ leads to Eq. (3) immediately. Why so simple transformation like (9) can change Dirac equation into a "superluminal" equation? It is because the $U_s$ in (9) is a nonhermitian matrix. While a unitary but hermitian matrix like that in (7) merely amounts to a representation transformation, a unitary but nonhermitian matrix is capable of changing the physical essence of equation. Eq.(11) can be expressed in Dirac representation as
$$
\begin{array}{l}
i\hbar \frac \partial {\partial t}\psi_s
=ic\hbar\vec{\alpha}\cdot \nabla \psi_s+\beta_s m_sc^2\psi_s
\end{array}\eqno{(12)}
$$ $$ \psi_s=\left(\begin{array}{c}\varphi_s\\ \chi_s\end{array}\right)
=U\left(\begin{array}{c}\xi_s\\ \eta_s\end{array}\right),\quad \beta_s=
\left(\begin{array}{cc}0&I \\-I&0\end{array}\right).\eqno{(13)} $$ Note that the
matrix $\beta_s$ and its counterpart in Eq. (11) are antihermitian.

On the other hand, if we directly set $m_0=im_s$ in Eq. (5) to get the "Dirac equation with imaginary mass" as
$$
\begin{array}{l}
i\hbar \frac \partial {\partial t}\psi^{(i)}
=ic\hbar\vec{\alpha}\cdot \nabla \psi^{(i)}+\beta_s^{(i)} m_sc^2\psi^{(i)}
\end{array},\eqno{(14)}
$$
then we have
$$
\psi^{(i)}=\left(\begin{array}{c}\varphi^{(i)}\\ \chi^{(i)}\end{array}\right),\quad
\beta_s^{(i)}= \left(\begin{array}{cc}iI&0 \\0&-iI\end{array}\right).\eqno{(15)}
$$
The fact that equation (14) is wrong can easily be seen by its plane-wave solution when $p\longrightarrow 0$, yielding a decoupling solution:
$$
\begin{array}{l}
\varphi^{(i)}\sim e^{-i(im_{s} t)}\sim e^{m_{s}t}\\
\chi^{(i)}\sim e^{-i(-im_{s} t)}\sim e^{-m_{s}t}
\end{array}\eqno{(16)}
$$ which show the violation of unitarity. By contrast, such kind of meaningless
solution is definitely excluded in Eq. (12). The sharp difference is stemming from the
following fact. Three kinds of Eqs. (5), (12) and (14), all respect the basic
symmetry: under the space-time inversion ($\vec{x}\longrightarrow -\vec{x}$,
$t\longrightarrow -t$) $^{[8]}$ $$ \varphi (-\vec{x},-t)\longrightarrow \chi
(\vec{x},t),\quad \chi (-\vec{x},-t) \longrightarrow \varphi (\vec{x},t),\eqno{(17)}
$$ the theory remains invariant (while a concrete solution of particle transforms into
that of antiparticle). However, we should consider a smaller symmetry---the
space-inversion ($\vec{x}\longrightarrow -\vec{x}$, $t\longrightarrow t$): $$ \xi
(-\vec{x},t)\longrightarrow \eta (\vec{x},t),\quad \eta(-\vec{x},t)\longrightarrow
\xi(\vec{x},t).\eqno{(18)} $$ Then we see that Eqs. (5) and (14) remain invariant
whereas Eq. (12) (or (11)) fails to do so. In other words, our new superluminal
equation reflects the maximum parity violation, a property exactly explaining the
permanent helicity of neutrino---while $\nu_{_L}$ and $\overline \nu_{_R}$ are
physically realized, $\nu_{_R}$ and $\overline \nu_{_L}$ are forbidden strictly---as
verified by experiments $^{[9,10]}$.

\section{Virtual particle as a superluminal particle}
\hskip\parindent It is no surprise to see solution like (16) when Eq. (15) contains an
imaginary mass $im_s$ explicitly. Actually, in quantum mechanics, we often endow an
imaginary part of mass $m$ to wave function for describing an unstable particle: $$
m\longrightarrow m-i\frac{\Gamma}{2},\quad \Gamma=\hbar/\tau, \quad\mid\psi\mid^2\sim
e^{-t/\tau}\eqno(19) $$ with $\tau$ being the decay lifetime. Of course, the unitarity
of wave function is destroyed.

Hence, it is hopeless to set $m_0=im_s$ directly in Dirac equation for describing a
stable superluminal particle. However, we often discuss the "virtual particle" in
covariant perturbation theory in the sense that its momentum $p$ and energy $E$ can
vary independently and so are not subjected to the constraint $E^2-p^2=m^2_0$. Both
$E^2>p^2$ and $E^2<p^2$ cases can occur. We might as well say that a virtual particle
could be superluminal but it is unstable too. A stable superluminal particle can only
be realized as $\nu_{_L}$ or $\overline\nu_{_R}$ ensured by the violation of parity
symmetry.

\section{Maximum parity violation displays the beauty of nature}
\hskip\parindent Usually, a symmetry, i.e., an invariance under some transformation
yields an affirmative guarantee for a theory and its violation seems to spoil the
validity of the theory. Now we see that for a meaningful discrete symmetry like
space-inversion, what happens in nature is either keeping intact ( for subluminal
particle ) or being violated to maximum ( for superluminal neutrino ). In the latter
case the violation must reach its maximum because the constraint of larger
symmetry---the space-time invariance---must be held at the same time. [In the first
line of Eq. (11), the coefficient of $m_s$ must be fixed as $(-1)$ to ensure the
invariance of transformation (17) together with the violation of transformation (18).]
We might as well look at the maximum violation of parity as some "antisymmetry" (
rather than "asymmetry" ) which also provides affirmative guarantee for the validity
of superluminal theory for neutrino. In fact, the normal consequence of violation of
hermitian property being the violation of unitarity ( as shown by (16)) is now recast
into a strange realization---of four would be unstable solutions for a same momentum,
two of them ($\nu_{_R}$ and $\overline\nu_{_L}$) are eliminated whereas other two
($\nu_{_L}$ and $\overline\nu_{_R}$) are stabilized.

\section{Summary and discussion}
\hskip\parindent
We can have three kinds of Dirac-type equation:
$$
\begin{array}{l}
i\hbar \frac \partial {\partial t}\psi
=ic\hbar\vec{\alpha}\cdot \nabla \psi+\beta mc^2\psi
\end{array}\eqno{(20)}
$$
as listed in the following table.

\begin{table}[ht]
\begin{center}
\renewcommand{\arraystretch}{1.5}
\small
\tabcolsep 4pt
\begin{tabular}{|c|c|c|c|}
\hline && Superluminal equation& Dirac equation with\\ property&Dirac equation &Eq.
(11), (12)& imaginary mass\\ & & & Eq. (14)\\ \hline $m$ (mass)&$m_0$ (rest
mass)&$m_s$ (proper mass)&$m_s$\\ \hline Dirac representation &$\beta=
\left(\begin{array}{cc}I&0
\\0&-I\end{array}\right)$ &$\beta_s= \left(\begin{array}{cc}0&I
\\-I&0\end{array}\right)$ &$\beta_s^{(i)}= \left(\begin{array}{cc}iI&0 \\0 & -iI
\end{array} \right)$ \\ $\psi=\left(\begin{array}{c}\varphi\\
\chi\end{array}\right),\quad \vec{\alpha}=\left(\begin{array}{cc}0 & \vec{\sigma} \\
\vec{\sigma}&0\end{array}\right)$ &hermitian matrix & antihermitian matrix &
antihermitian matrix \\ \hline Weyl representation&$\beta= \left(\begin{array}{cc}0&I
\\I&0\end{array}\right)$ &$\beta_s= \left(\begin{array}{cc}0&-I
\\I&0\end{array}\right)$ &$\beta_s^{(i)}= \left(\begin{array}{cc}0&iI
\\iI&0\end{array}\right)$ \\ $\psi=\left(\begin{array}{c}\xi\\
\eta\end{array}\right),\quad \vec{\alpha}=\left(\begin{array}{cc}\vec{\sigma}&0 \\
0&-\vec{\sigma}\end{array}\right)$ & hermitian matrix & antihermitian matrix &
antihermitian matrix \\ \hline hermitian property of theory&yes&no&no\\ \hline
unitarity of theory&yes&yes&no\\ \hline invariance under & & &\\ space-time
inversion&yes&yes&yes\\
 (basic symmetry)& & &\\
\hline
invariance under space&yes&no&yes\\
inversion (parity)&&&\\
\hline
&& describing possibly the & unstable superluminal \\
physical meaning&sublumial particles & (stable) superluminal& particle without\\
&(electron, etc.)& neutrino $\nu_{_L}$ and $\overline\nu_{_R}$& physical meaning\\
& & (with $\nu_{_R}$ and $\overline\nu_{_L}$ forbidden)& in reality\\
\hline
\end{tabular}\end{center}\end{table}

Finally, two remarks are in order:

(a) According to our present understanding, no boson but fermion can be superluminal
as long as the parity symmetry is violated to maximum. Most likely, one kind of known
particles, the neutrino, is just a superluminal particle, a tachyon with spin 1/2.

(b)The physical implication of nonhermitian transformation $U_s$ amounts to an extra
phase difference $\pi$/2 between $\xi_s$ and $\eta_s$ in comparison with that in
subluminal case. Hence we see onceagain that it is the phase which plays a dominant
role in quantum mechanics.

(c)The origin ( or mechanism ) of mass generation has been remained in physics as a
mystery for hundred years. It is time to say something now. According to detailed
analysis in Ref. [6], the finite and changeable mass of a fermion is not due to one
kind of excitation which varies continuously inside but a manifestation of coherent
cancellation effect between two fields rotating with opposite helicities implicitly,
either one of them can be excited infinitely in essence. It is precisely this
fantastic effect explains the amazing result that a tachyon's energy (mass) approaches
zero when its velocity $u$ increases to infinity (as shown in Eq. (4)).

Acknowledgements: The author wishes to thank Dr. Tsao Chang for bringing the
superluminal problem of neutrino to his attention and relevant discussions.

\end{document}